\documentclass[reprint,longbibliography,amsmath,amssymb,showpacs,aps,prl]{revtex4-1}
\usepackage{graphicx}
\usepackage[colorlinks,linkcolor=blue,citecolor=blue,urlcolor=blue]{hyperref}
\usepackage{multirow}
\usepackage{subfigure}

\begin{document}

\title{Anisotropy-induced Extrinsic Chirality and Chiral Discrimination of Surface Plasmon Polaritons}

\author{Qiang Zhang}
\author{Junqing Li}
\email{jqli@hit.edu.cn}
\author{Rui Zhao}
\author{Xingguang Liu}
\author{Gebeyehu Dirbeba}

\affiliation{School of Physics, Harbin Institute of Technology, 92 Western Dazhi, Harbin 150001, China}

\begin{abstract}
We present the characteristics of a simple waveguiding structure constructed by anisotropic birefringent crystal-metal-chiral medium, anisotropic-metal-chiral in short, and reveal the chiral-dependent dispersion and propagation properties of the surface plasmon polaritons (SPPs). We demonstrate its remarkable discrimination ability to the magnitude and sign of both the real and imaginary part of the chirality parameter. The anisotropy plays a key role in such performance and shows tuneable ability in enantiomeric discrimination even when the chirality parameter is complex-valued. Most importantly, the physical origin of chiral discrimination stems from the extrinsic chirality of the system, which arises from the mutual orientation of the SPPs and the optical axis. Moreover, we also clarify the fundamental physics behind the chiral discriminating behaviour by associating the intrinsic quantum spin Hall effect (QSHE) of the SPPs with the electromagnetic field analysis. This structure does not rely on complicated fabrication but provides the opportunity of on-chip surface-sensitive biosensing. We anticipate that our work will stimulate intensive research to investigate the anisotropy-induced chiral sensing techniques in plasmonic platforms.
\end{abstract}

\maketitle

Chirality refers to the symmetry property of an object that is not congruent with its mirror image. Manifestations of chirality exist ubiquitously in nature from the DNA molecule to human hands and maelstrom in the ocean and spiral galaxy in the universe. In optics, chiral
medium exhibits optical activity, also known as circular birefringence (CB), which describes the ability to rotate the polarization state of light and circular dichroism (CD), which measures the differential transmission of circularly polarized waves. Based on these effects, many optical schemes have been proposed for chiral sensing and discrimination, that is a terminology for the method to differentiate chiral enantiomers (the chiral object and its mirror image). Chiral discrimination is especially important for pharmaceutical applications as nearly 50$\%$ of the drugs are chiral and among which most of their chiral enantiomeric couterparts are toxic\cite{C6AY02015A,PMC}. However, discrimination of chiral objects via optical measurements by CB and CD are generally challenging because the chiral-dependent signals are inherently weak. Engineering complex structures \cite{Liu2009,Ren2012,Cai2014,Mario,Lodahl2017,doi:10.1021/acsphotonics.8b00270}, such as plasmonic metamaterials, would amplify the detected signals but it heavily relies on sophisticated complex nanofabrication procedures.

Here we propose an anisotropic-metal-chiral waveguiding structure and address its ability to unambiguously detect both the real and imaginary part of the chirality parameter $\kappa$. The introduction of a structure with a plethora of novel materials has been reported as enhancing its optical performance. We consider anisotropic crystal as a good candidate because it is convincing that anisotropy would provide the conventional waveguides with brand new properties, as many works have discovered novel phenomena such as a new class of hybridized SPPs\cite{1069973,4101,PhysRevB.75.235420,PhysRevB.81.085426,PhysRevB.81.115335,doi:10.1063/1.2908920,doi:10.1063/1.3541653,Zhou_2015,PhysRevB.85.085442,doi:10.1063/1.4773877,plasmonics2013,PhysRevA.100.033809} and tunable photonic bound states in the continuum (BICs)\cite{Gomis-Bresco2017,PhysRevA.98.063826,PhysRevA.100.053819,Mukherjee:19} in waveguiding structures that contain anisotropic crystals. Meanwhile, the introduction of chirality to convention plasmonic waveguides also leads to interesting physics such as enhancement of SPPs propagation and hybridized SPPs\cite{Mi:14,Zhang:16,doi:10.1063/1.4982158}. However, in these works, enantiomeric chirality parameters that contain a pair of opposite handedness would result in the same effect to the system and thereby failing in chiral discrimination. A recent work shows that sub-millidegree angle-resolved chiral surface plasmon resonance (CHISPR) scheme can detect the absolute
chirality (handedness and magnitude)\cite{10.1021/acsphotonics.9b00137}, which provides another proof that chiral plasmonics has prominent application in chiral sensing. It would be interesting and inspiring to investigate the problem of what will happen if both the anisotropy and the chirality are involved in a plasmonic waveguide? 

We theoretically answer this question by demonstrating that the proposed anisotropic-metal-chiral structure enables the detection of the magnitude and sign of both the real and imaginary part of the chirality parameter $\kappa$. And for the first time the dispersion relation equation and the corresponding physical pictures are presented in such a system. We show the fundamental characteristics of the SPPs in the structure including the theoretical derivation and analysis of the anisotropic-chiral SPPs. Most importantly we discuss the fundamental physics behind our findings.

\begin{figure}[htbp!]
\centering
\includegraphics{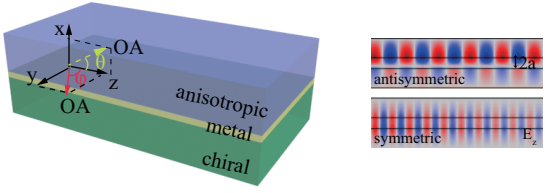}
\caption{Schematic of anisotropic-metal-chiral waveguiding structure with the optical axis (OA) in the $yz$ or $xz$ plane. The system supports antisymmetric mode and symmetric mode.}{\label{fig1}}
\end{figure}

We restrict ourselves to the simplest waveguiding structure fabricated with uniaxially anisotropic crystal, metal and chiral medium; the optical axis (OA) of the crystal is set to lie in the $yz$ or $xz$ plane that forms an angle $\varphi$ or $\theta$ with respect to the wave propagation direction $z$, respectively, as shown in Fig.~\ref{fig1}. The more complicated case when the OA has projections in both $yz$ or $xz$ plane is not the focus of this work for simplicity as it is intuitively a problem of the summation of the two separate cases above. We also would not consider the excitation of SPPs for the same reason.

In a chiral medium: ${\bf D}=\epsilon_c{\bf E}-i\kappa\sqrt{\mu_0\epsilon_0}{\bf H}$, ${\bf B}=\mu_c{\bf H}+i\kappa\sqrt{\mu_0\epsilon_0}{\bf E}$, where $\kappa$, $\epsilon_c$ and $\mu_c$ are the chirality strength, the permittivity and permeability of the chiral material, respectively. $k_\pm=(\sqrt{{\epsilon_c\mu_c}/{\epsilon_0\mu_0}}\pm\kappa)\omega/c$ are two eigen-wavenumbers corresponding to the right-handed circularly polarized (RCP) wave and left-handed circularly polarized (LCP) wave, respectively. In the bulk material, Maxwell's equations have a general form
$\vec{k}\times(\vec{k}\times\vec{E})+\hat{\epsilon}k_0^2\vec{E}=0$. For the anisotropic medium, we only consider when the optical axis lies in the $yz$ plane\cite{doi:10.1063/1.3541653}(when it lies in the $xz$ plane, no coupling of SPPs from two interfaces exist, see Supplemental Material), 
\begin{equation}
\hat{\epsilon}=
\begin{bmatrix}
\begin{smallmatrix}
\epsilon_x' & 0 & 0\\
0 & \epsilon_z' \sin^2\varphi+\epsilon_y' \cos^2\varphi & (\epsilon_y'-\epsilon_z') \sin\varphi\cos\varphi\\
0 & (\epsilon_y'-\epsilon_z') \sin\varphi\cos\varphi & \epsilon_z' \cos^2\varphi+\epsilon_y' \sin^2\varphi\\
\end{smallmatrix}
\end{bmatrix}\label{m}
\end{equation}
with $\vec{k}=(i\alpha,0,\beta)$, where $\beta$ is the propagation constant and $\alpha$ is the decay constant of the SPPs, and $\epsilon_z'=\epsilon_e$, $\epsilon_x'=\epsilon_y'=\epsilon_o$ ($\epsilon_o<\epsilon_e$), representing the relative permittivity for the extraodinary and ordinary wave, respectively. After some deduction procedure (see Supplemental Material), we get two independent decay constants for the ordinary wave and extraordinary wave in the anisotropic medium $\alpha_o=\sqrt{\beta^2-\epsilon_ok_0^2}$ and $\alpha_e=\sqrt{\beta^2(\epsilon_0\sin^2\varphi+\epsilon_e\cos^2\varphi)/\epsilon_o-\epsilon_ek_0^2}$, respectively. Applying boundary conditions yields the dispersion relation for the SPPs:
\begin{widetext}
\begin{equation}
\begin{vmatrix}
\begin{smallmatrix}
\alpha_o\cos\varphi & \sin\varphi & -e^{2\alpha_ma} & -1 & 0 & 0 & 0 & 0\\
\alpha_o\sin\varphi & \frac{\alpha^2_o\cos\varphi}{\epsilon_ok_0^2} & 0 & 0 & -\frac{\alpha_m}{\omega\epsilon_m}e^{2\alpha_ma} & \frac{\alpha_m}{\omega\epsilon_m} & 0 & 0\\
\frac{\epsilon_ok_0^2\sin\varphi}{\omega\mu_0} & \frac{\alpha_e\cos\varphi}{\omega\mu_0} & 0 & 0 & e^{2\alpha_ma} & 1 & 0 & 0\\
\alpha_o^2\cos\varphi & \alpha_e\sin\varphi & \epsilon_me^{2\alpha_ma} & -\epsilon_m & 0 & 0 & 0 & 0\\
0 & 0 & 1 & e^{2\alpha_ma} & 0 & 0 & 1 & -1\\
0 & 0 & 0 & 0 &  \frac{\alpha_m}{\omega\epsilon_m} & -\frac{\alpha_m}{\omega\epsilon_m}e^{2\alpha_ma} & \frac{\alpha_+}{k_+} & \frac{\alpha_-}{k_-}\\
0 & 0 & 0 & 0 & \eta & \eta e^{2\alpha_ma} & 1 & 1\\
0 & 0 & \frac{\alpha_m\eta}{\omega\mu_0} & -\frac{\alpha_m\eta}{\omega\mu_0}e^{2\alpha_ma} &  0 & 0 & \frac{\alpha_+}{k_+} & -\frac{\alpha_-}{k_-}\\
\end{smallmatrix}
\end{vmatrix}\label{matrix1}
=0,
\end{equation}
\end{widetext}
where $\eta=\sqrt{\mu_c/\epsilon_c}$ is the wave impedance of the chiral medium, $\epsilon_m$ the permittivity of the metal, $\alpha_m$ the SPPs decay constants in the metal (Drude model, in this work we use gold and the parameters can be found in Ref.\cite{Zhang:16}) and $\alpha_{\pm}$ the decay constants in the chiral medium for RCP and LCP with $\alpha_{\pm}=\beta^2-k_{\pm}^2$.

\begin{figure*}[htbp!]
\centering
\includegraphics{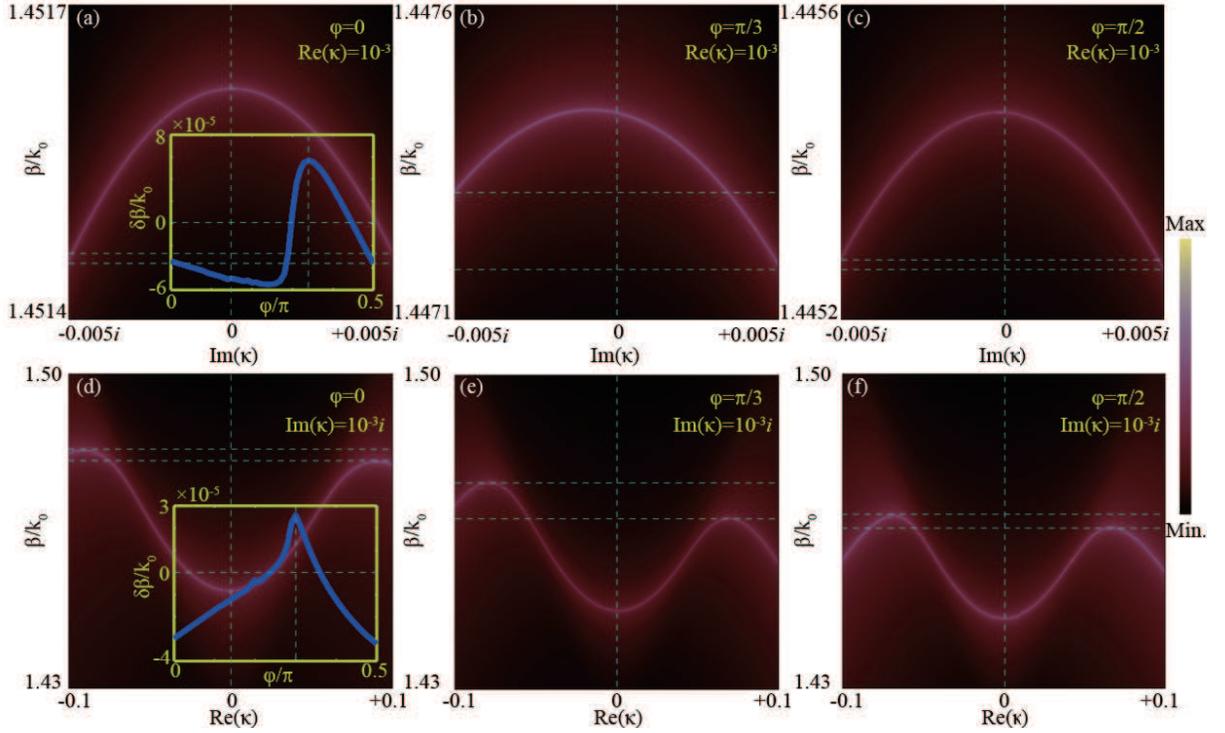}
\caption{Sensitivity of the propagation constant $\beta$ to the real and imaginary part of complex-valued $\kappa$ for three typical optical axis orientations $\varphi=0$, $\varphi=\pi/3$ and $\varphi=\pi/2$.(a-c) Density plot of the normalized propagation constant $\beta/k_0$ as a function of the imaginary part of chirality Im($\kappa$)=$-0.1i\sim+0.1i$, when Re($\kappa$)=$10^{-3}$. (d-f) Density plot of $\beta/k_0$ as a function of the real part of chirality Re($\kappa$)=$-0.1\sim+0.1$, when Im($\kappa$)=$10^{-3}i$. The insets show the optimal $\varphi$ occurs near $\pi/3$, discussion about dependence of optimal $\varphi$ on material parameter choice is in the Supplemental Material.}{\label{fig2}}
\end{figure*}
\begin{figure}[htbp!]
\centering
\includegraphics{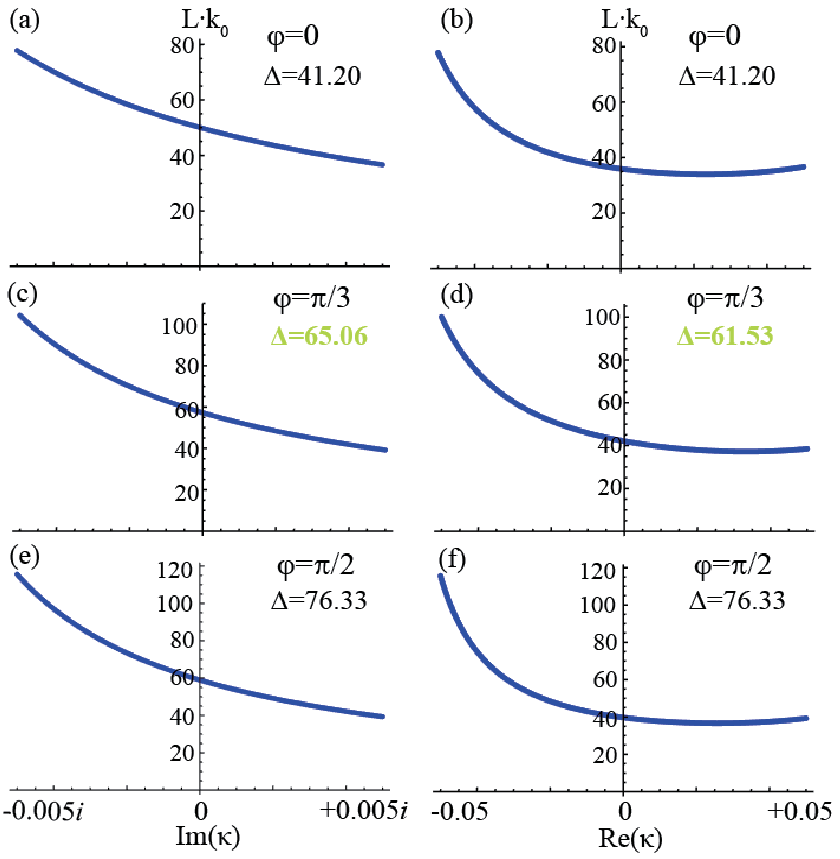}
\caption{Sensitivity of the propagation distance $L$ to the real and imaginary part of $\kappa$ where $\Delta=L\cdot k_0\mid_{Im(\kappa)=-0.005i}-L\cdot k_0\mid_{Im(\kappa)=+0.005i}$ and $\Delta=L\cdot k_0\mid_{Re(\kappa)=-0.05}-L\cdot k_0\mid_{Re(\kappa)=+0.05}$ respectively in the left and right columns. Left column: scan of Im($\kappa$) with Re($\kappa$) fixed at +0.05 for (a) $\varphi=0$,  (c) $\varphi=\pi/3$ and (e) $\varphi=\pi/3$. Right column: scan of Re($\kappa$) with Im($\kappa$) fixed at $+0.005i$ for (b) $\varphi=0$, (d) $\varphi=\pi/3$, and (f) $\varphi=\pi/2$.}{\label{fig3}}
\end{figure}
We stress here that the choice of material parameters such as the permittivity would not affect the main conclusions we get in this paper. Throughout this paper we set $2a=20$ nm, $\epsilon_c=2+0.02i$, $\epsilon_o=1.44$, $\epsilon_e=2.89$ and a representative wavelength at $\lambda=1550$ nm for its broad applications in optics and communications. The dispersion diagrams for various central film thickness and the discussion on material parameters are shown in the Supplemental Material.
In Fig.~\ref{fig2} we show the dependence of the propagation constant ($\beta$) at $\lambda=1550$ nm on the real and imaginary part of the chirality parameter $\kappa$, with different optical axis angles ($\varphi$) taken into consideration. The diagrams are drawn by the density plot of the inverse determinant of the boundary condition matrix (Equation (\ref{matrix1})) of the system. This is a standard method to obtain a plot whose maxima correspond to the dispersion branches\cite{HassaniGangaraj:19}. Our discussion with respect to $\varphi$ is restricted in the range of 0 to $\pi/2$ for simplicity since the result would be reversed for $\pi/2<\varphi<\pi$ considering the mirror symmetry.

To avoid violating the passivity, we ensure that Im($n_c+\kappa$)$>0$, so $\vert$Im($\kappa$)$\vert$ should be smaller than 0.007. In Fig.~\ref{fig2}a, b and c, we vary Im($\kappa$) between -0.005$i$ and +0.005$i$ and fix Re($\kappa$) at $10^{-3}$, where we can see that high symmetry is maintained with subtle difference for $\beta$ at $\pm$Im($\kappa$) as we scan Im($\kappa$) when $\varphi=0$ and $\varphi=\pi/2$. While an apparent differential $\beta$ is observed for $\pm$Im($\kappa$) when $\varphi=\pi/3$. Meanwhile, in Fig.~\ref{fig2}d, e and f, we vary Re($\kappa$) between -0.1 and +0.1 to better illustrate the differtial propagation constant and fix Im($\kappa$) at $10^{-3}i$ where an obvious increase of difference is also obtained for $\varphi=\pi/3$, see Fig.~\ref{fig2}e. More discussions of multiple combinations of Re($\kappa$) and Im($\kappa$) can be found in the Supplemental Material.

In fact, $\beta$ can not be measeured directly, what measurable is the propagation length $L$ ($L=1/2\vert$Im$(\beta)\vert$) detected by near-field imaging\cite{Marquart:05,Jahng:15}. Fig.~\ref{fig3} shows the propagation length as functions of the real and imaginary part of $\kappa$. To clarify the relationship between $L$ and $\kappa$, we calculate $L\cdot k_0$ instead of $L$, where we choose $\lambda=1550$ nm in our calculation. We scan $\kappa$ among $\pm0.05\pm0.005i$ and investigate the dependence of $L$ on its magnitude and sign. We also study the enantiomeric discrimination of the complex-valued chirality of the system. 
Intriguingly, when $0<\varphi<\pi/2$, $L\mid_{+\kappa}\neq L\mid_{-\kappa}$ as the illustrations with respect to $\varphi=\pi/3$ in Fig.~\ref{fig3}c and Fig.~\ref{fig3}d obviously demonstrate that $\delta L=(L\cdot k_0\mid_{+0.05-0.005i}-L\cdot k_0\mid_{+0.05+0.005i})/k0= 65.06/k0 = 16.06 \mu m$ and $\delta L=(L\cdot k_0\mid_{-0.05+0.005i}-L\cdot k_0\mid_{+0.05+0.005i})/k0 = 61.53/k0 = 15.18 \mu m$. This indicates a differential propagation length of 880 nm occurs for $\kappa=+0.05-0.005i$ and $\kappa=-0.05+0.005i$. Therefore, when the optical axis has an orientation $0<\varphi<\pi/2$, the anisotropic-chiral SPPs can discriminate chiral enantiomers even when the chirality parameter is complex valued as the propagation length of the SPPs are different for $\pm\kappa$. However, 
such a behavior vanishes as implied by the same $\Delta$ in Fig.~\ref{fig3}a, b and c, d when $\varphi=0$ and $\pi/2$.  

\begin{figure}
\centering
\includegraphics{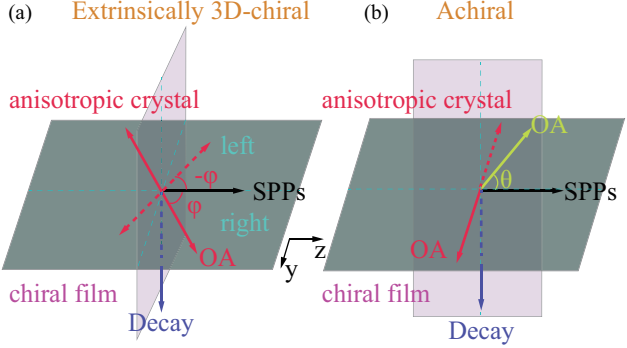}
\caption{Origin of chiral sensing: mirror asymmetry of the experimental arrangement. According to the SPP decay direction towards the chiral medium (blue), the SPP momentum (black) and the OA orientation (red), the system is (a) extrinsically 3D-chiral and right-handed when $0<\varphi<\pi/2$ and left-handed when $-\pi/2<\varphi<0$ and (b) achiral when $\varphi=0$, $\varphi=\pi/2$ and when OA (yellow) lies in the $xz$ plane.}{\label{fig4}}
\end{figure}

Such a simple structure offers us the opportunity of on-chip bio-sensing with easy fabrication and surface detection. We argue that the chiral sensing ability of the proposed structure originates from the extrinsic 3D-chirality of the system, which highly resembles the extrinsic 3D-chirality in 2D metamaterials\cite{PhysRevLett.102.113902,doi:10.1063/1.4954033}. Here, chirality is drawn extrinsically from the mutual orientation of the SPP decay direction pointing to the chiral medium, the SPP propagation direction and the optical axis as schematically illustrated by Fig.~\ref{fig4}.

In Fig.~\ref{fig4}a, when the optical axis lies in the waveguide plane and forms an azimuth angle $\varphi$ with respect to the SPP propagation direction, the extrinsic 3D-chirality is readily seen by considering the experimental arrangement of the three colored solid arrows. Since the SPP direction is fixed along $+z$ and we focus on the SPPs that close to the chiral medium which constantly decay towards the chiral medium, the system is mirror asymmetric and the chirality of is solely determined by the orientation of the optical axis, namely right-handed when $0<\varphi<\pi/2$ and left-handed when $-\pi/2<\varphi<0$ (equivalent to $\pi/2<\varphi<\pi$). In particular, as shown in Fig.~\ref{fig4}b: 1) when $\varphi=0$ that is the OA superimposes on the SPP propagation, the system loses chirality; 2) when $\varphi=\pi/2$ that is the OA lies orthoganal to the SPP propagation, the system lacks chirality because of failure to determine the system's handedness as OA lies in $+y$ and $-y$ means the same thing; 3) when the optical axis lies perpendicular to the waveguide plane and forms an inclination angle $\theta$ with respect to the SPP propagation direction, the three colored arrows will be co-planar thereby losing chirality. These facts essentially explain the origin of the chiral sensing effect and why when the OA lies at $\varphi=0$ and $\varphi=\pi/2$ the system can not discriminate chiral enantiomers.

\begin{figure}
\centering
\includegraphics{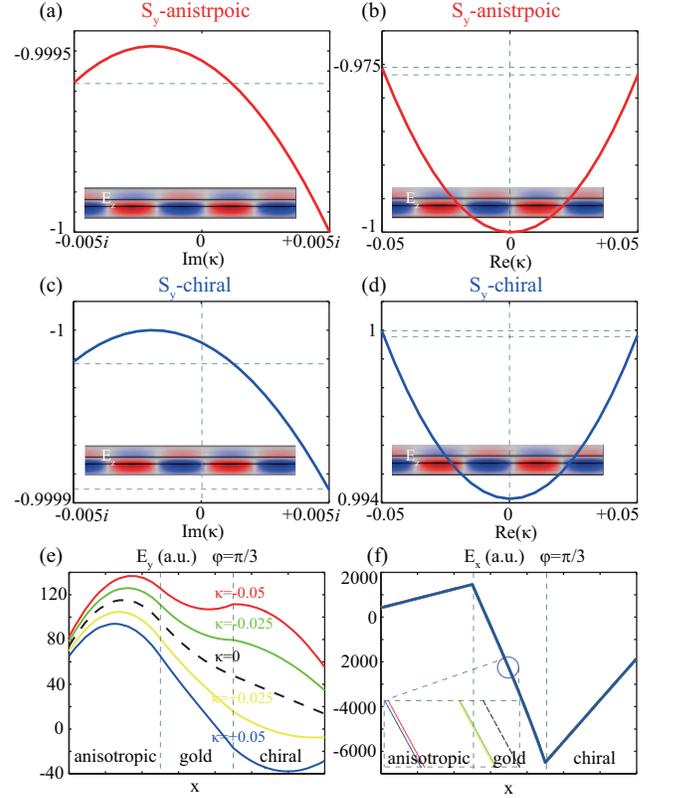}
\caption{Transverse SAM ($S_y$) distributions in the system as functions of the real and imaginary part of $\kappa$, when $\varphi=\pi/3$. (a, c) $S_y$-Im($\kappa$) and (b, d) $S_y$-Re($\kappa$) in the anisotropic crystal and chiral medium, respectively. (e) Transverse $E_y$ component distribution and (f) longitudinal $E_z$ component distribution shows chiral-dependent and -independent behaviors, respectively.}{\label{fig5}}
\end{figure}

It has been shown that the transverse spin angular momentum (SAM) carried by SPPs is a manifestation of the intrinsic quantum spin Hall effect (QSHE) of light\cite{Bliokh1448,PhysRevLett.103.100401,PhysRevA.85.061801,Bliokh20151,Bliokh,Alizadeh,doi:10.1021/acsphotonics.5b00574}. On the basis of this fundamental effect, we study numerically the impact of anisotropy and chirality on distribution of the transverse SAM and electric field in the system. Fig.~\ref{fig5} describes the chiral-dependent transverse SAM $S_y$ when $\varphi=\pi/3$, where $S_y=i\epsilon_0/(4\omega\mu_c)(\bf{E}\times\bf{E}^*)+ i\mu_0/(4\omega\epsilon_c)(\bf{H}\times\bf{H}^*)$\cite{PhysRevLett.114.063901,4941539}. For simplicity we set $\kappa$ to be pure imaginary-valued and real-valued. It is clear that the unequivalent transverse SAM at $\pm$Im($\kappa$) exist (but not exist when $\varphi=0$ and $\varphi=\pi/2$, see Suppplmental Material). The opposite signs of transverse SAM distribution in the two media are due to spin-locking property\cite{4941539} of SPPs. This proves the validity of our anlytical results previously shown. Fig.~\ref{fig3}e reveals a distinct behavior contrary to isotropic systems that the transverse $E_y$ component emerges even when $\kappa=0$ if $0<\varphi<\pi/2$. The introduction of anisotropy leads to asymmetric $E_y$ distribution for $\pm$Im($\kappa$) and $\pm$Re($\kappa$) and thereby forming a substantial change of TM to TE conversion that results in the differential propagation length $\delta\beta$. Interestingly, Fig.~\ref{fig3}f presents that the longitudinal electric field component depends neglegibly on the sign and magnitude of $\kappa$ since the cuves overlap with each other. Therefore, it is not the change of refractive index due to the orientation of optical axis nor the magnitude change of chirality that causes the chiral discrimination. Notably, the existence of hybridized SPPs due to anisotropy is the fundamental reason for the discrimination of the chirality.

It is necessary to stress here that $S_y$ essentially is a quantity that can be measured indirectly. A particle in the surface plasmonic field experiences an anomalous lateral optical force perpendicular to the direction of the momentum of the SPPs\cite{Bliokh,Wang,4941539}. Such lateral force is proportional to the transverse SAM thus it can be harnessed for manipulation of small particles which in turn can be regarded as a feedback for chiral sensing of the thin (subwavelength) chiral films.

In summary, the proposed simple system allows unambiguous sensing and detection of the magnitude and handedness of both the real and imaginary part of the chirality parameter $\kappa$. This phenomenon is due to the extrinsic chirality, which arises from the mutual orientation of the SPPs and the optical axis. Specifically, when the optical axis lies at azimuth angle $\varphi=0$ or $\pi/2$, the SPPs can detect the magnitude and sign of both the real and imaginary part of the chirality parameter $\kappa$ \emph{but can not} discriminate chiral enantiomers; when $0<\varphi<\pi/2$, however, the SPPs can detect the magnitude and sign of both the real and imaginary part of the chirality parameter $\kappa$ \emph{as well as} discriminate chiral enantiomers even when $\kappa$ is complex-valued. The chiral sensitivity of the proposed system can be realized by rotating the waveguide so that the optical axis orientation determines the measurement of chiral-dependent differential propagation length of the SPPs. Such a simple waveguiding structure offers an theoretical instruction to chiral plasmonic device design and can be of great interest in chiral-biosensing applications.

\textbf{Acknowledgements}

This work was supported by the National Natural Science Foundation of China (NSFC) under grant number 60977032 and the Program for Innovation Research of Science  of Harbin Institute of Technology (PIRS-HIT), China (Grant No. T201407).

\bibliography{apssamp}

\end{document}